\documentclass[twocolumn,author-numerical]{revtex4-1}
\usepackage{amsmath,amsthm,epsfig,graphicx,amssymb}
\usepackage{dcolumn}
\usepackage{bm}
\usepackage{axodraw4j}
\begin{document}
\setcitestyle{super}
\title{SU(4) based classification of four-level systems and their semiclassical solution}
\author{Surajit Sen}
\email[Corresponding author:\quad]{ssen55@yahoo.com}
\author{Helal Ahmed}
\affiliation{Physics Department, Guru Charan College, Silchar 788004, India}
\begin{abstract}
We present a systematic method to classify the four-level system using $SU(4)$ symmetry as the basis group. It is shown that this symmetry allows three dipole transitions which eventually leads to six possible configurations of the four-level system. Using a dressed atom approach, the semi-classical version of each configuration is exactly solved under rotating wave approximation and the symmetry of the Rabi oscillation among various models is studied and its implication is discussed.
\end{abstract}
\maketitle
\section{\label{1x} Introduction}
\par
Last year we observe the centenary of the epoch-making discovery of the `atomic orbit' postulated by Bohr which is the central idea to understand the origin of the atomic spectra~\cite{Bohr1913,Bohr1914}. These orbits are indeed stationary energy levels with well-defined quantum numbers and plays fundamental role in the development of modern quantum physics. In recent years, advent of laser technology and high-Q cavity leads to the experimental realization of so called Rydberg atom where the energy levels with longer lifetime and large dipole moment can be prepared. Because of the large value of the dipole moment, the allowed dipole transition between the levels can be maneuvered in a selective and controlled manner. The system with two well-defined levels interacting with the quantized cavity field is  known as Jaynes-Cummings model which is exactly solvable under the rotating wave approximation~\cite{Jaynes1963}. The pretext of this fully quantized version of two-level system was, however, set by the semiclassical two-level system originally proposed to formulate the theoretical basis of the nuclear magnetic resonance~\cite{Rabi1937,Gerry2000}. In the semiclassical two-level system the interacting oscillatory electromagnetic field is treated classically while the atom is treated quantum mechanically. An immediate extension of the two-level system is the three-level system which is associated with a rich class of coherent phenomena, namely, two photon coherence, double resonance process, three-level super-radiance, resonance Raman scattering, population trapping, tri-level echoes, STIRAP, quantum jump, quantum zeno effect, electromagnetically induced transparency (EIT), etc~\cite{Yoo1985,Ref2012}.
It is therefore worth investigating the four-level system which may be associated more phenomena of light-matter interaction uncharted so far. Apart from that, there is another reason of studying such model. In recent times, the manipulation and control of quantum mechanical systems using multiple electromagnetic fields is an area of intense research in the parlance of quantized control theory ~\cite{}. Thus developing a systematic and rigorous theory of a semiclassical and quantized four-level system may provide some new insights into the area of atomic, molecular and optical physics.
\par
Although the four-level system is instrumental in understanding the population inversion scenario by the optical pumping method, its use in context with quantum optical models is, however, not very large. Some studies along this direction includes the four-level EIT effect~\cite{Li2008,Qi2010,Joshi2005}, dynamics of the pulse propagation through coherently prepared four-level system~\cite{Paspalakis2002}, modeling the qubit-induced micro-switching~\cite{Ham2000}, Rabi oscillation in cascade four-level system~\cite{Nath2008a} etc. All these studies deal with the model Hamiltonians which are proposed phenomenologically and therefore lacks proper understanding of the selection rules of allowed transition among various levels. In a recent investigation we have shown that the three-level system can be successfully classified using $SU(3)$ as the basis group~\cite{Nath2008b,Sen2012}. It is therefore interesting to look for the straight forward but non-trivial extension of the treatment to study the four-level system where the $SU(4)$ group plays key role to identify the possible allowed dipole transitions. In particular we show that the appropriate choice of $SU(4)$ basis leads to a systematic classification of the four-level systems.
\par
The primary objective of the paper is to discuss the possible classification of the four-level system using the $SU(4)$ group as the basis group and then to look for the semi-classical solution of the model Hamiltonians under rotating wave approximation. To achieve this goal, the remaining Sections of the paper are organized as follows; in Section-II we discuss the essential properties of $SU(4)$ group necessary to formulate the model Hamiltonians of all possible four-level configurations. In Section-III, we develop the methodology to solve these models using Bose-Pascos matrix, a generalized six-parameter Euler matrix in four dimension~\cite{Bose1980}. The numerical studies are presented in Section IV to compare the Rabi oscillation of all semiclassical models. In the concluding Section we summarize the main results of the paper and discuss the outlook.
\section{\label{2x} Four-level system and $SU(4)$ group}
\par
The generic Hamiltonian of an arbitrary four-level system which allows all possible transitions is represented by the hermitian matrix,
\begin{equation}
H = \left[ {\begin{array}{*{20}{c}}
{{\Delta _{44}}}&{{h_{43}}}&{{h_{42}}}&{{h_{41}}}\label{one}\\
{{h_{43}}}&{{\Delta _{33}}}&{{h_{32}}}&{{h_{31}}}\\
{{h_{42}}}&{{h_{32}}}&{{\Delta _{22}}}&{{h_{21}}}\\
{{h_{41}}}&{{h_{31}}}&{{h_{21}}}&{{\Delta _{11}}}
\end{array}} \right],
\end{equation}
\noindent
where ${h_{ij}}$  $(i,j = 1,2,3,4)$  be the matrix element and $\Delta_{ij}$ is the detuning of the applied trichromatic field which vanishes at resonance.
Of these four levels, two levels are involved in each transition and we have $\frac{4!}{(4-2)!2!}=6$ possible configurations of the four-levels system with three possible dipole transitions shown in Fig.1-6.
For Model-I we note that the non-vanishing terms $h_{41}\ne 0,h_{32}\ne 0,h_{21}\ne 0$ in Eq.~(\ref{one}) which correspond to the allowed dipole transitions $4 \leftrightarrow 1, 3 \leftrightarrow 2$ and $2 \leftrightarrow 1$, respectively, while the remaining three transitions are forbidden. Proceeding in the similar
way all six possible Hamiltonians of the four-level system can be built up and Table-I illustrates the requirement of their construction:
\begingroup
\begin{table}[h]
\caption{}
\begin{ruledtabular}
\begin{tabular}{lll}
Model & Forbidden transition & Allowed transition\\
\hline
I & $h_{43}=0, h_{42}=0, h_{31}=0 $ & $h_{41}\ne 0,h_{32}\ne 0,h_{21}\ne 0$ \\
II & $h_{41}=0, h_{42}=0, h_{32}=0 $ & $h_{43}\ne 0,h_{31}\ne 0,h_{21}\ne 0$ \\
III & $h_{42}=0, h_{41}=0, h_{31}=0 $ & $h_{43}\ne 0,h_{32}\ne 0,h_{21}\ne 0$\\
IV & $h_{42}=0, h_{32}=0, h_{31}=0 $ & $h_{43}\ne 0,h_{41}\ne 0,h_{21}\ne 0$ \\
V & $h_{41}=0, h_{32}=0, h_{31}=0 $ & $h_{43}\ne 0,h_{42}\ne 0,h_{21}\ne 0$ \\
VI & $h_{31}=0, h_{42}=0, h_{21}=0 $ & $h_{43}\ne 0,h_{41}\ne 0,h_{32}\ne 0$\\
\end{tabular}
\end{ruledtabular}
\end{table}
\endgroup
\par
In order to construct the Hamiltonians quantitatively using $SU(4)$ as the basis group, we shall briefly recall its properties ~\cite{Greiner1994}.
\par
The $SU(4)$ group is described by fifteen $\lambda_i$ ($i=1,2,...15$) matrices, which follow the following commutation, anti-commutation and normalization relations,
\begin{subequations}
\begin{eqnarray}\label{two}
[{\lambda _i},{\lambda _j}] &=& 2i{f_{ijk}}{\lambda _k}, \quad \quad
\{ {\lambda _i},{\lambda _j}\}  = {\delta _{ij}}I + 2{d_{ijk}}{\lambda _k},\\
Tr[{\lambda _i}] &=& 0, \qquad\qquad \quad Tr[{\lambda _i}{\lambda _j}] = 2{\delta _{ijk}},
\end{eqnarray}
\end{subequations}
\noindent
respectively. Here, $d_{ijk}$ and $f_{ijk}$ ($i,j,k = 1,2,..15$) are the completely symmetric and completely anti-symmetric structure functions which characterize the SU(4) group are defined as,
\begin{subequations}
\begin{eqnarray}\label{two}
{f_{ijk}} &=& \frac{1}{{4i}}Tr([{\lambda _i}{\lambda _j}]{\lambda _k}),\\
{d_{ijk}} &=& \frac{1}{4}Tr(\{ {\lambda _i}{\lambda _j}\} {\lambda _k}).
\end{eqnarray}
\end{subequations}
It is customary to define the $SU(4)$ shift operators as the linear combination of the $\lambda_i$ matrices,
\begin{widetext}
\begin{subequations}
\begin{eqnarray}\label{four}
    {T_ \pm } &=& \frac{1}{2}({\lambda _1} \pm i{\lambda _2}),\quad\quad{U_ \pm } = \frac{1}{2}({\lambda _6} \pm i{\lambda _7}),\quad\quad{V_ \pm } = \frac{1}{2}({\lambda _4} \pm i{\lambda _5}),\\
    {W_ \pm } &=& \frac{1}{2}({\lambda _9} \pm i{\lambda _{10}}),\quad\quad{X_ \pm } = \frac{1}{2}({\lambda _{11}} \pm i{\lambda _{12}}),\quad\quad{Z_ \pm } =
   \frac{1}{2}({\lambda _{13}} \pm i{\lambda _{15}}), \\
      {T_3} &=& {\lambda _3},\quad\quad {U_3} = \frac{1}{2}(\sqrt 3 {\lambda _8} - {\lambda _3}),\quad\quad
   {V_3} = \frac{1}{2}(\sqrt 3 {\lambda _8} + {\lambda _3}),\\
    {X_3} & = &  - \frac{1}{2}{\lambda _3} + \frac{1}{{2\sqrt 3 }}{\lambda _8} + \sqrt {\frac{2}{3}} {\lambda _{15}},\quad\quad {W_3} = \frac{1}{2}{\lambda _3} + \frac{1}{{2\sqrt 3 }}{\lambda _8} + \sqrt {\frac{2}{3}} {\lambda _{15}}, \\
    {Z_3} &=&  - \frac{1}{{\sqrt 3 }}{\lambda _8} + \sqrt {\frac{2}{3}} {\lambda _{15}},
\end{eqnarray}
\end{subequations}
\end{widetext}
which follow the closed algebra of $SU(4)$ group ~\cite{Greiner2008,Nath2008a}. Having defining the $SU(4)$ shift vectors, we now proceed to develop the Hamiltonians of all four-level configurations.
\begin{figure}[ht]
\begin{flushleft}
\begin{picture}(300,86)(0,0)

\Line(100,100)(150,100)
\Text(180,100)[]{$E_4,|4>$}

\Line(100,70)(150,70)
\Text(180,70)[]{$E_3,|3>$}

\Line(100,40)(150,40)
\Text(180,40)[]{$E_2,|2>$}
\Text(90,40)[]{$\Delta_{32}$}

\Line(100,10)(150,10)
\Text(180,10)[]{$\Delta_{21},E_1,|1>$}
\Text(90,10)[]{$\Delta_{41}$}

\ArrowLine(110,100)(110,12)
\ArrowLine(125,70)(125,43)
\ArrowLine(140,40)(140,14)

\Text(110,105)[]{$\omega _{41}$}
\Text(125,75)[]{$\omega _{32}$}
\Text(140,45)[]{$\omega _{21}$}

\DashCurve{(120,43)(130,43)}{1.0}
\DashCurve{(130,14)(150,14)}{1.0}
\DashCurve{(100,12)(125,12)}{1.0}

\end{picture} \\
\begin{flushleft}
Fig.1: The energies of the four levels of Model-I are $E_1=-\omega_1-\omega_2$, $E_2=\omega_2-\omega_3$, $E_3=\omega_3$ and $E_4=\omega_1$, respectively
\end{flushleft}
\end{flushleft}
\vspace {.25cm}
\begin{flushleft}
\begin{picture}(300,86)(0,0)

\Line(100,100)(150,100)
\Text(180,100)[]{$\Delta_{43},E_4,|4>$}

\Line(100,70)(150,70)
\Text(180,70)[]{$E_3,|3>$}

\Line(100,40)(150,40)
\Text(180,40)[]{$E_2,|2>$}

\Line(100,10)(150,10)
\Text(180,10)[]{$\Delta_{21},E_1,|1>$}
\Text(90,10)[]{$\Delta_{31}$}

\ArrowLine(140,100)(140,73)
\ArrowLine(110,70)(110,12)
\ArrowLine(140,40)(140,14)

\Text(140,105)[]{$\omega _{43}$}
\Text(110,77)[]{$\omega _{31}$}
\Text(140,45)[]{$\omega _{21}$}

\DashCurve{(130,14)(150,14)}{1.0}
\DashCurve{(130,73)(150,73)}{1.0}
\DashCurve{(100,12)(125,12)}{1.0}
\end{picture} \\
\begin{flushleft}
Fig.2: The energies of the four levels of Model-II are $E_1=-\omega_1-\omega_3$, $E_2=\omega_1$, $E_3=-\frac{\omega_2}{2}+\omega_3$, $E_4=\frac{\omega_2}{2}$ and respectively.
\end{flushleft}
\end{flushleft}
\vspace {.25cm}
\begin{flushleft}
\begin{picture}(300,86)(0,0)

\Line(100,100)(150,100)
\Text(180,100)[]{$E_4,|4>$}

\Line(100,70)(150,70)
\Text(180,70)[]{$E_3,|3>$}
\Text(90,70)[]{$\Delta_{43}$}

\Line(100,40)(150,40)
\Text(180,40)[]{$\Delta_{32},E_2,|2>$}

\Line(100,10)(150,10)
\Text(180,10)[]{$\Delta_{21},E_1,|1>$}

\ArrowLine(110,100)(110,73)
\ArrowLine(125,70)(125,43)
\ArrowLine(140,40)(140,14)

\Text(110,105)[]{$\omega _{43}$}
\Text(125,77)[]{$\omega _{32}$}
\Text(142,45)[]{$\omega _{21}$}

\DashCurve{(100,73)(125,73)}{1.0}
\DashCurve{(115,43)(135,43)}{1.0}
\DashCurve{(130,14)(150,14)}{1.0}
\end{picture} \\
\begin{flushleft}
Fig.3: The energies of the four levels of Model-III are $E_1=-\omega_1$, $E_2=\omega_1-\omega_3$, $E_3=-\omega_2+\omega_3$ and  $E_4=\omega_2$, respectively
\end{flushleft}
\end{flushleft}
\end{figure}
\section{\label{3x} The models}
To obtain the Hamiltonian of Model-I in the $SU(4)$ basis, we write the non-vanishing terms in Eq.(1) in the following form,
\begin{equation}\label{five}
{H_I(t)} = \left[ {\begin{array}{*{20}{c}}
{{E_4}}&0&0&{{\kappa _{41}}{e^{ - i{\omega _{41}}t}}}\\
0&{{E_3}}&{{\kappa _{32}}{e^{ - i{\omega _{32}}t}}}&0\\
0&{{\kappa _{32}}{e^{i{\omega _{32}}t}}}&{E_2}&{{\kappa _{21}}{e^{ - i{\omega _{21}}t}}}\\
{{\kappa _{41}}{e^{i{\omega _{41}}t}}}&0&{{\kappa _{21}}{e^{i{\omega _{21}}t}}}&{E_1}
\end{array}} \right],
\end{equation}
where, $\omega_{ab}$ and $\kappa_{ab}$ ($a,b=1,2,3,4, a\neq b$) be the frequencies of the applied tri-chromatic field and the coupling parameters, respectively. If we define the energy levels to be, $E_4=\omega_1$, $E_3=\omega_3$, $E_2=\omega_2-\omega_3$ and $E_1=-\omega_1-\omega_2$,
\begin{figure}
\begin{center}
\begin{picture}(300,86)(0,0)

\Line(100,100)(150,100)
\Text(180,100)[]{$E_4,|4>$}

\Line(100,70)(150,70)
\Text(180,70)[]{$\Delta_{43},E_3,|3>$}

\Line(100,40)(150,40)
\Text(180,40)[]{$E_2,|2>$}

\Line(100,10)(150,10)
\Text(180,10)[]{$\Delta_{43},E_1,|1>$}
\Text(90,10)[]{$\Delta_{41}$}

\ArrowLine(110,100)(110,12)
\ArrowLine(140,100)(140,73)
\ArrowLine(140,40)(140,14)

\Text(110,105)[]{$\omega _{41}$}
\Text(140,105)[]{$\omega _{43}$}
\Text(140,45)[]{$\omega _{21}$}

\DashCurve{(130,14)(150,14)}{1.0}
\DashCurve{(100,12)(125,12)}{1.0}
\DashCurve{(130,73)(150,73)}{1.0}

\end{picture} \\
\begin{flushleft}
Fig.4: The energies of the four levels of Model-IV are $E_1=-\omega_1-\omega_3$, $E_2=\omega_1$, $E_3=-\frac{\omega_2}{2}$, $E_4=\frac{\omega_2}{2}+\omega_3$, respectively,
\end{flushleft}
\end{center}
\vspace{.25cm}
\begin{center}
\begin{picture}(300,86)(0,0)

\Line(100,100)(150,100)
\Text(180,100)[]{$E_4,|4>$}

\Line(100,70)(150,70)
\Text(180,70)[]{$\Delta_{43},E_3,|3>$}

\Line(100,40)(150,40)
\Text(180,40)[]{$E_2,|2>$}
\Text(90,40)[]{$\Delta_{42}$}

\Line(100,10)(150,10)
\Text(180,10)[]{$\Delta_{21},E_1,|1>$}

\Text(110,105)[]{$\omega _{42}$}
\Text(140,105)[]{$\omega _{43}$}
\Text(140,45)[]{$\omega _{21}$}

\ArrowLine(110,100)(110,43)
\ArrowLine(140,100)(140,73)
\ArrowLine(140,40)(140,14)

\DashCurve{(130,14)(150,14)}{1.0}
\DashCurve{(100,43)(125,43)}{1.0}
\DashCurve{(130,73)(150,73)}{1.0}

\end{picture} \\
\begin{flushleft}
Fig.5: The energies of the four levels of Model-V are $E_1=-\omega_1$, $E_2=\omega_1-\omega_3$, $E_3=-\omega_2$ and  $E_4=\omega_2+\omega_3$, respectively
\end{flushleft}
\end{center}
\vspace{.25cm}
\begin{center}
\begin{picture}(300,86)(0,0)

\Line(100,100)(150,100)
\Text(180,100)[]{$E_4,|4>$}

\Line(100,70)(150,70)
\Text(180,70)[]{$\Delta_{43},E_3,|3>$}

\Line(100,40)(150,40)
\Text(180,40)[]{$\Delta_{32},E_2,|2>$}

\Line(100,10)(150,10)
\Text(180,10)[]{$E_1,|1>$}
\Text(90,10)[]{$\Delta_{41}$}

\ArrowLine(110,100)(110,14)
\ArrowLine(125,70)(125,43)
\ArrowLine(140,100)(140,73)

\Text(110,105)[]{$\omega _{41}$}
\Text(140,105)[]{$\omega _{43}$}
\Text(125,75)[]{$\omega _{32}$}

\DashCurve{(120,43)(130,43)}{1.0}
\DashCurve{(100,14)(125,14)}{1.0}
\DashCurve{(130,73)(150,73)}{1.0}

\end{picture} \\
\end{center}
\begin{flushleft}
Fig.6: The energies of the four levels of Model-VI are $E_1=-\omega_1$, $E_2=-\omega_3$, $E_3=-\frac{\omega_2}{2}+\omega_3$ and $E_4=\frac{\omega_2}{2}+\omega_1$, respectively
\end{flushleft}
\end{figure}
respectively, Eq.(5) can be equivalently expressed as,
\begin{eqnarray}\label{six}
{H_I(t)} &=& {\omega _1}{W_3} + {\omega _2}{Z_3} + {\omega _3}{U_3} + {\kappa _{41}}{W_ + }{e^{ - i{\omega _{41}}t}} \\ \nonumber
 &+&{\kappa _{21}}{Z_ + }{e^{ - i{\omega _{21}}t}} + {\kappa _{32}}{U_ + }{e^{ - i{\omega _{32}}t}} + h.c. .
\end{eqnarray}
This is precisely the Hamiltonian of Model-I in terms of shift operators where only three operators are involved.
The construction of the remaining Hamiltonians of other models involves judicious combination of the shift operators and we have the Hamiltonians,
\begin{eqnarray}\label{seven}
{H_{II}(t)} &=& {\omega _1}{Z_1} + {\omega _2}{T_3} + {\omega _3}{X_3} + {\kappa _{21}}{Z_ + }{e^{ - i{\omega _{21}}t}} \\ \nonumber
&+& {\kappa _{43}}{T_ + }{e^{ - i{\omega _{43}}t}} + {\kappa _{31}}{X_ + }{e^{ - i{\omega _{31}}t}} + h.c.,
\end{eqnarray}
for Model-II,
\begin{eqnarray}\label{eight}
{H_{III}(t)} &=& {\omega _1}{Z_3} + {\omega _2}{T_3} + {\omega _3}{U_3} + \kappa _{21}{Z_ + }{e^{ - i{\omega _{21}}t}} \\ \nonumber
&+& \kappa _{43}{T_ + }{e^{ - i{\omega _{43}}t}} + \kappa _{32}{U_ + }{e^{ - i{\omega _{32}}t}}+h.c.,
\end{eqnarray}
for Model-III,
\begin{eqnarray}\label{nine}
{H_{IV}(t)} &=& {\omega _1}{Z_3} + {\omega _2}{T_3} + {\omega _3}{W_3} + {\kappa _{21}}{Z_ + }{e^{ - i{\omega _{21}}t}} \\ \nonumber
&+& {\kappa _{43}}{T_ + }{e^{ - i{\omega _{43}}t}} + {\kappa _{41}}{W_ + }{e^{ - i{\omega _{41}}t}} + h.c.,
\end{eqnarray}
for Model-IV,
\begin{eqnarray}\label{ten}
{H_{V}(t)} &=& {\omega _1}{Z_3} + {\omega _2}{T_3} + {\omega _3}{V_3} + \kappa _{21}{Z_ + }{e^{ - i{\omega _{21}}t}} \\ \nonumber
&+&\kappa _{43}{T_ + }{e^{ - i{\omega _{43}}t}} + \kappa _{42}{V_ + }{e^{ - i{\omega _{42}}t}}+h.c.
\end{eqnarray}
for Model-V,
\begin{eqnarray}\label{eleven}
{H_{VI}(t)} &=& {\omega _1}{W_3} + {\omega _2}{T_3} + {\omega _3}{U_3} +\kappa _{41}{W_+}{e^{ - i{\omega _{41}}t}}\\ \nonumber
&+&\kappa _{43}{T_+}{e^{ - i{\omega _{43}}t}}+\kappa _{32}{U_+}{e^{ - i{\omega _{32}}t}}+h.c.,
\end{eqnarray}
for Model-VI, respectively.

In presence of interaction, let the solution of the four-level system of Model-I in Eq.~(\ref{six}) is given by,
\begin{equation}\label{twelve}
\Psi_I(t) = {C_1}(t)\left| 1 \right.\rangle  + {C_2}(t)\left| 2 \right.\rangle  + {C_3}(t)\left| 3 \right.\rangle  + {C_4}(t)\left| 4 \right.\rangle,
\end{equation}
where ${C_1}(t),{C_2}(t),{C_3}(t)$ and ${C_4}(t)$  are the normalized time-independent amplitudes which are to be calculated with the basis states given by,
\begin{equation}\label{thirteen}
\left| 1 \right.\rangle  = \left[ {\begin{array}{*{20}{c}}
0\\
0\\
0\\
1
\end{array}} \right], \left| 2 \right.\rangle  = \left[ {\begin{array}{*{20}{c}}
0\\
0\\
1\\
0
\end{array}} \right], \left| 3 \right.\rangle  = \left[ {\begin{array}{*{20}{c}}
0\\
1\\
0\\
0
\end{array}} \right], \left| 4 \right.\rangle  = \left[ {\begin{array}{*{20}{c}}
1\\
0\\
0\\
0
\end{array}} \right].
\end{equation}
The wave function in Eq.~(\ref{twelve}) obeys the time-dependent Schr$\ddot{o}$dinger equation,\\
\begin{equation}\label{fourteen}
i\hbar \frac{{\partial \Psi_I }}{{\partial t}} = H_I(t)\Psi_I.
\end{equation}
The Hamiltonian can be made time-independent by the transformation,
\begin{equation}\label{fifteen}
\tilde H_I(0) =  - i\hbar {U_I^\dag }\dot U_I + {U_I^\dag }H_I(t)U_I,
\end{equation}
and the basis in which is the energy is diagonalized can be obtained by a unitary transformation,
\begin{equation}\label{sixteen}
\tilde \Psi_I(t) = {U_I}(t)\Psi_I(0).
\end{equation}
For Model-I, the unitary operator is defined as,
\begin{subequations}\label{seventeen}
\begin{eqnarray}
U_{I}(t) &=& exp[({\frac{i}{4}(2{\omega _{21}} +{\omega _{32}} - 3{\omega _{41}})}{W_3}\\ \nonumber
&+&{\frac{i}{2}(-2{\omega _{21}}-{\omega _{32}}+{\omega _{41}})}{Z_3}\\ \nonumber
&+&\frac{i}{4}(-2{\omega _{21}}-3{\omega _{32}}+{\omega _{41}}){U_3})t].
\end{eqnarray}
\end{subequations}
Using Eq.~(\ref{seventeen}), the time-independent Hamiltonian in Eq.~(\ref{fifteen}) takes the following form,
\begin{equation}\label{eighteen}
{\tilde H_I(0)} =\left[ {\begin{array}{*{20}{c}}
{\Delta_{44}^I}&0&0&{{\kappa _{41}}}\\
0&{\Delta_{33}^I}&{{\kappa _{32}}}&0\\
0&{{\kappa _{32}}}&{\Delta_{22}^I}&{{\kappa _{21}}}\\
{{\kappa _{41}}}&0&{{\kappa _{21}}}&{\Delta_{11}^I}
\end{array}} \right],
\end{equation}
where, the diagonal terms are given by,
\begin{subequations}\label{nineteen}
\begin{eqnarray}
\Delta_{44}^I
&=& \frac{1}{4}(4\omega_{1}+2\omega_{21}+\omega_{32}-\omega_{41}),\\
\Delta_{33}^I
&=& \frac{1}{4}(4\omega_{3}-2\omega_{21}-3\omega_{32}+\omega_{41}),\\
\Delta_{22}^I
&=& \frac{1}{4}(4\omega_{2}-2\omega_{21}-4\omega_{3}+\omega_{32}+\omega_{41}),\\
\Delta_{11}^I
&=& \frac{1}{4}(-4\omega_{1}-4\omega_{2}+2\omega_{21}+\omega_{32}+\omega_{41}).
\end{eqnarray}
\end{subequations}
In Eq.~(\ref{nineteen}) the diagonal terms can be expressed in terms linear combination of the detuning,
\begin{subequations}
\begin{eqnarray}\label{twenty}
\Delta_{44}^I
&=& \frac{1}{4}( - 2\Delta _{21}^I - \Delta _{32}^I + 3\Delta _{41}^I),\\
\Delta_{33}^I
&=& \frac{1}{4}(2\Delta _{21}^I + 3\Delta _{32}^I - \Delta _{41}^I),\\
\Delta_{22}^I
&=& \frac{1}{4}(2\Delta _{21}^I - \Delta _{32}^I - \Delta _{41}^I),\\
\Delta_{11}^I
&=&  - \frac{1}{4}(2\Delta _{21}^I + \Delta _{32}^I + \Delta _{41}^I),
\end{eqnarray}
\end{subequations}
respectively, where the detuning from the applied field $\omega_{ab}$ are given by
\begin{subequations}
\begin{eqnarray}\label{twentyone}
\Delta _{21}^I &=& (2{\omega _2} - {\omega _3} + {\omega _1}) - {\omega _{21}}, \\
\Delta _{32}^I &=& ( - {\omega _2} + 2{\omega_3}) - {\omega_{32}},\\
\Delta _{41}^I &=& ({\omega_2} + 2{\omega_1}) - {\omega_{41}},
\end{eqnarray}
\end{subequations}
respectively, depicted if Fig.1. The derivation of the unitary operators and the time-independent Hamiltonians of the remaining models are similar and we quote the results in Appendix.
\par
At resonance ($\Delta _{44}^I=0$, $\Delta _{33}^I=0$, $\Delta _{22}^I=0$ and $\Delta _{11}^I=0$), the time-dependent probability amplitudes of the four-levels are given by,
\vspace{1.5cm}
\begin{widetext}
\begin{equation}\label{twentytwo}
\left[ {\begin{array}{*{20}{c}}
{{C_4}(t)}\\
{{C_3}(t)}\\
{{C_2}(t)}\\
{{C_1}(t)}
\end{array}} \right] =
T_\alpha ^{ - 1}\left[ {\begin{array}{*{20}{c}}
{{e^{ - i{\Lambda _4}t}}}&0&0&0\\
0&{{e^{ - i{\Lambda _3}t}}}&0&0\\
0&0&{{e^{ - i{\Lambda _2}t}}}&0\\
0&0&0&{{e^{ - i{\Lambda _1}t}}}
\end{array}} \right]{T_\alpha }\left[ {\begin{array}{*{20}{c}}
{{C_4}(0)}\\
{{C_3}(0)}\\
{{C_2}(0)}\\
{{C_1}(0)}
\end{array}} \right],
\end{equation}
\end{widetext}
where $\Lambda_i$ be the resonant eigenvalues of the time-independent Hamiltonian Eq.~(\ref{eighteen}) and $T_\alpha$ be the six parameter Bose-Pascos orthogonal matrix given by ~\cite{Nath2008a,Bose1980},
\begin{equation}\label{twentythree}
{T_\alpha } = \left[ {\begin{array}{*{20}{c}}
{{\alpha _{11}}}&{{\alpha _{12}}}&{{\alpha _{13}}}&{{\alpha _{14}}}\\
{\alpha {}_{21}}&{{\alpha _{22}}}&{{\alpha _{23}}}&{{\alpha _{24}}}\\
{{\alpha _{31}}}&{\alpha {}_{32}}&{\alpha {}_{33}}&{{\alpha _{34}}}\\
{{\alpha _{41}}}&{{\alpha _{42}}}&{{\alpha _{43}}}&{\alpha {}_{44}}
\end{array}} \right].
\end{equation}
The elements in Eq.~(\ref{twentythree}) reads,
\begin{equation}
\begin{array}{l}\label{twentyfour}
{\alpha _{11}} = {c_1}{c_5} + {s_1}{s_3}{s_4}{s_5}\\
{\alpha _{12}} = {c_1}{s_5}{s_6} + {s_1}{c_3}{c_6} + {s_1}{s_3}{s_4}{c_5}{s_6}\\
{\alpha _{13}} = {s_1}{s_3}{c_4}\\
{\alpha _{14}} =  - {c_1}{s_5}{s_6} - {s_1}{c_3}{s_6} + {s_1}{s_3}{s_4}{c_5}{s_6}\\
{\alpha _{21}} =  - {s_1}{c_2}{c_5} + ({c_1}{c_2}{s_3} - {s_2}{c_3}){s_4}{s_5}\\
{\alpha _{22}} = {s_1}{c_2}{s_5}{s_6} + {c_1}{c_2}{c_3} + {s_2}{s_3}){c_6} + ({c_1}{c_2}{s_3} \\
\qquad - {s_2}{c_3}){s_4}{c_5}{s_6}\\
{\alpha _{23}} = ({c_1}{c_2}{s_3} - {s_2}{c_3}){c_4}\\
{\alpha _{24}} = {s_1}{c_2}{s_5}{c_6} - ({c_1}{c_2}{c_3} + {s_2}{s_3}){c_6} + ({c_1}{c_2}{s_3} \\
\qquad - {s_2}{c_3}){s_4}{c_5}{s_6}\\
{\alpha _{31}} =  - {s_1}{s_2}{c_5} + ({c_1}{s_2}{s_3} + {c_2}{c_3}){s_4}\\
{\alpha _{32}} = {s_1}{s_2}{s_5}{s_6} + ({c_1}{c_2}{c_3} - {c_2}{s_3}){c_6} + ({c_1}{s_2}{s_3} \\
\qquad + {c_2}{c_3}){s_4}{c_5}{s_6}\\
{\alpha _{33}} = ({c_1}{s_2}{s_3} + {c_2}{c_3}){c_4}\\
{\alpha _{34}} = {s_1}{s_2}{s_5}{c_6} - ({c_1}{s_2}{c_3} - {c_2}{s_3}){s_6} + ({c_1}{s_2}{s_3} \\
 \qquad + {c_2}{c_3}){s_4}{c_5}{s_6}\\
{\alpha _{41}} = {c_4}{s_5}\\
{\alpha _{42}} = {c_4}{c_5}{s_6}\\
{\alpha _{43}} =  - {s_4}\\
{\alpha _{44}} = {c_4}{c_5}{c_6}\\
\end{array}
\end{equation}
where, ${s_i} = sin{\theta _i}$  and ${c_i} = cos{\theta _i}$ $(i=1,2,3,4,5,6)$. In the next Section we proceed to discuss the Rabi oscillation of various  levels numerically for with specific initial conditions.
\section{\label{5x} Numerical solutions}
To find the amplitudes of a given model we consider four possible initial conditions; Case-I: $C_1(0)=1, C_2(0)=0, C_3(0)=0, C_4(0)=0$, i.e., when the system is in the lowest state designated by level-1 (blue), Case-II: $C_1(0)=0, C_2(0)=1, C_3(0)=0, C_4(0)=0$, i.e., when the system is in the level-2 (green), Case-III: $C_1(0)=0, C_2(0)=0, C_3(0)=1, C_4(0)=0$, i.e., when the system is in the level-3 (red),
\begin{figure}[ht]
\includegraphics[width=8.25cm,height=8.25cm,keepaspectratio]{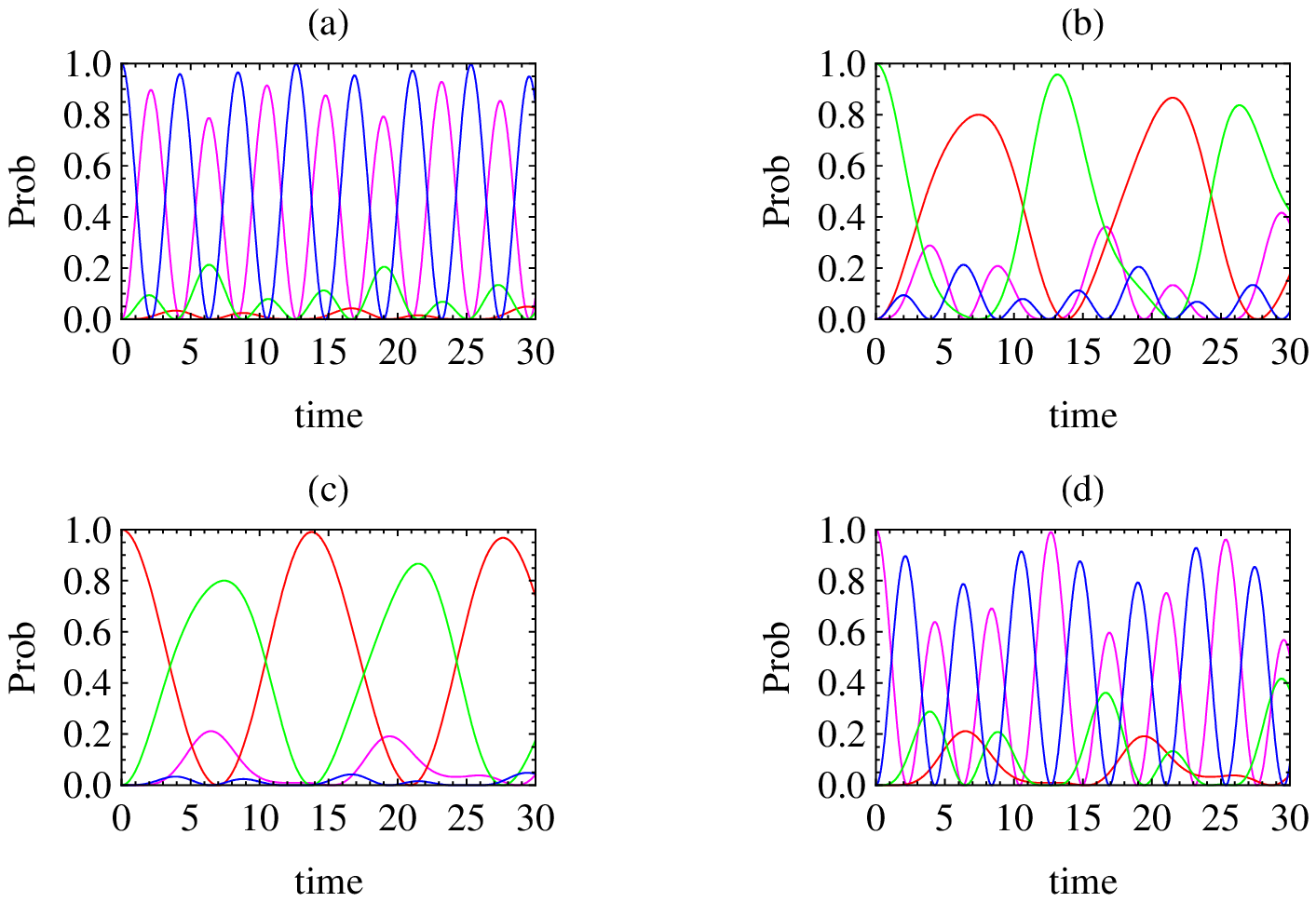}
\begin{flushleft}
Fig.7: Rabi oscillation of Model-I for Case-I (a), Case-II (b), Case-III (c) and Case-IV (d).
\end{flushleft}
\end{figure}
\begin{figure}[ht]
\includegraphics[width=8.25cm,height=8.25cm,keepaspectratio]{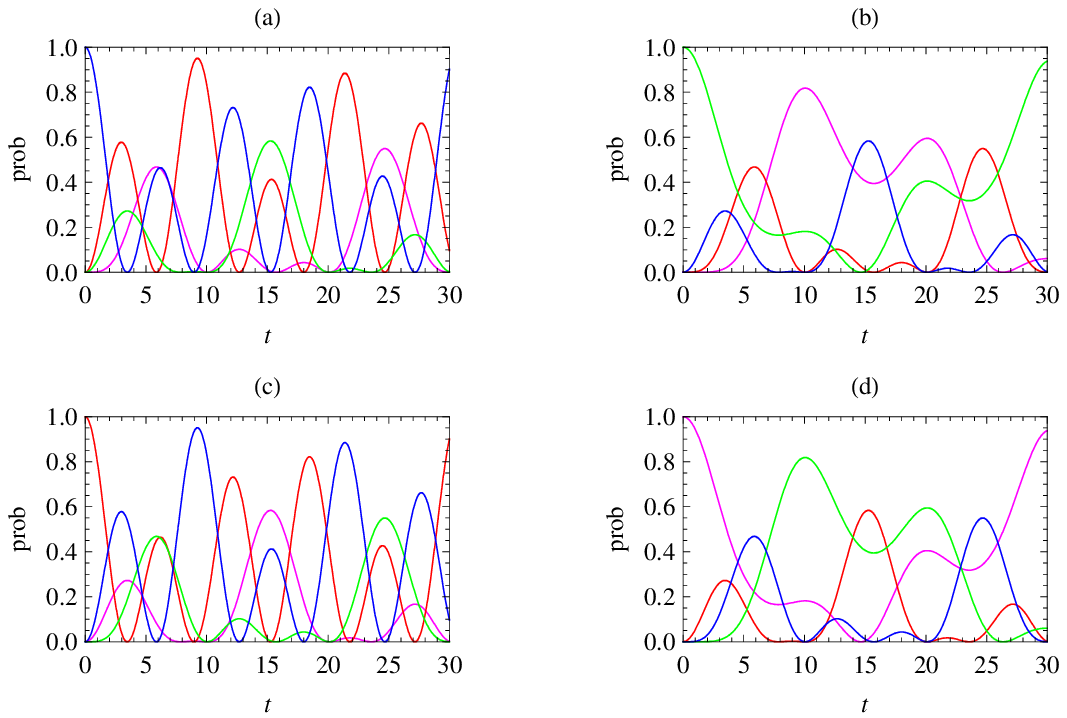}
\begin{flushleft}
Fig.8: Rabi oscillation of Model-II for Case-I (a), Case-II (b), Case-III (c) and Case-IV (d).
\end{flushleft}
\end{figure}
\begin{figure}[ht]
\includegraphics[width=8.25cm,height=8.25cm,keepaspectratio]{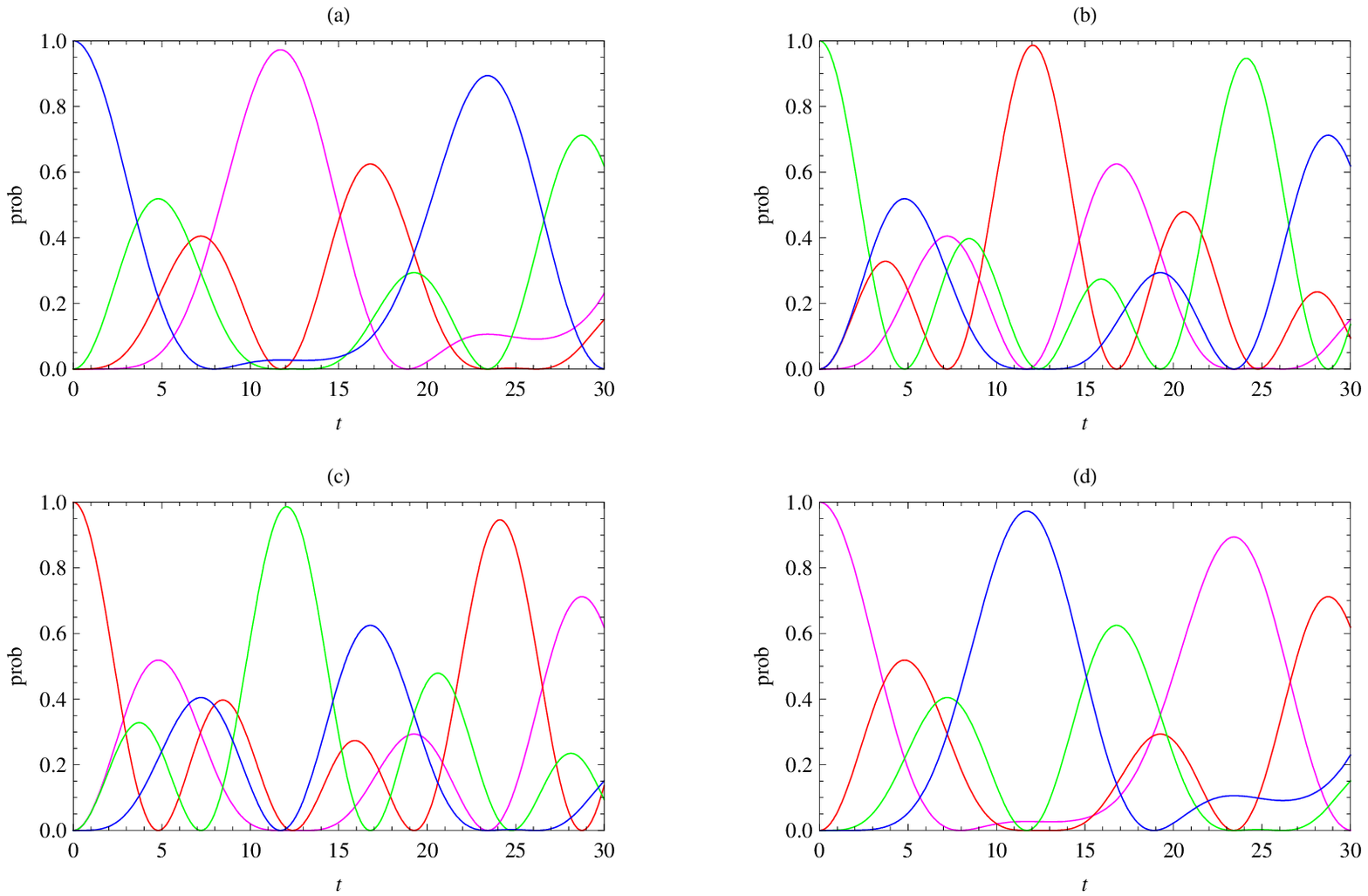}
\begin{flushleft}
Fig.9: Rabi oscillation of Model-III for Case-I (a), Case-II (b), Case-III (c) and Case-IV (d). The oscillation pattern of
(a) is similar to (c) and (b) is similar to (d), respectively.
\end{flushleft}
\end{figure}
\begin{figure}[ht]
\includegraphics[width=8.25cm,height=8.25cm,keepaspectratio]{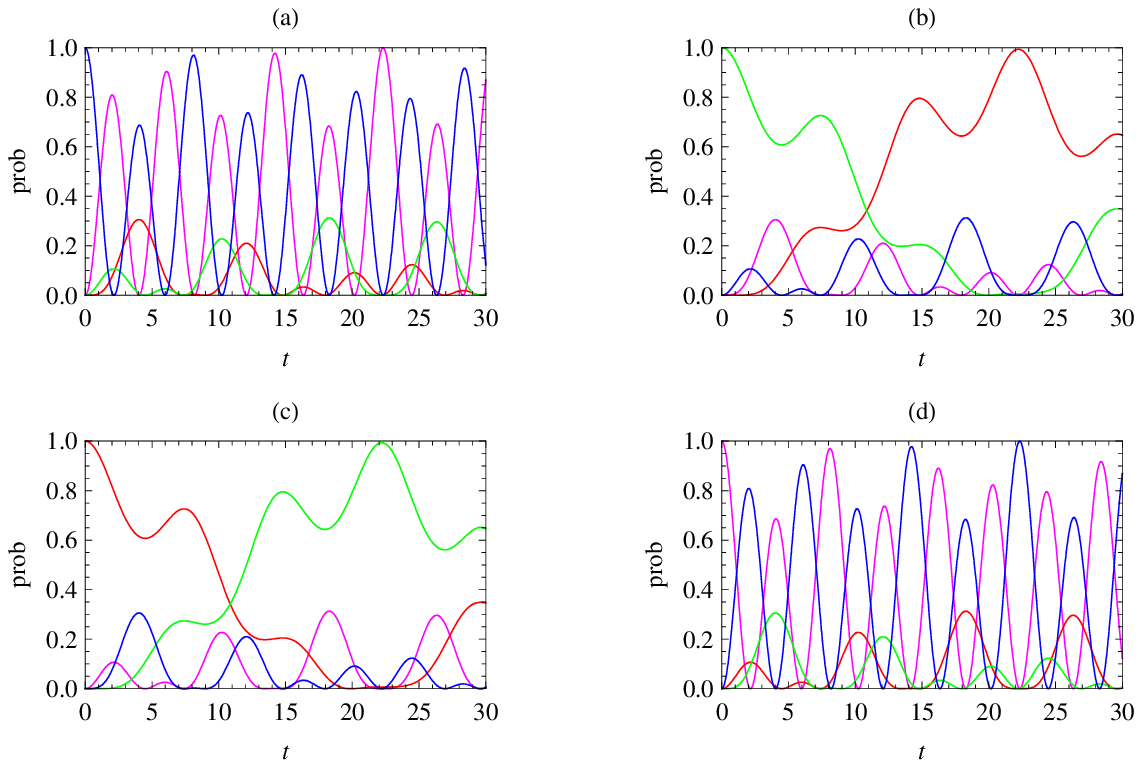}
\begin{flushleft}
Fig.10: Rabi oscillation of Model-IV for Case-I (a), Case-II (b), Case-III (c) and Case-IV (d) with $\omega_{43}=\omega_{32}=\omega_{21}=.4 $.
\end{flushleft}
\end{figure}
\begin{figure}[ht]
\includegraphics[width=8.25cm,height=8.25cm,keepaspectratio]{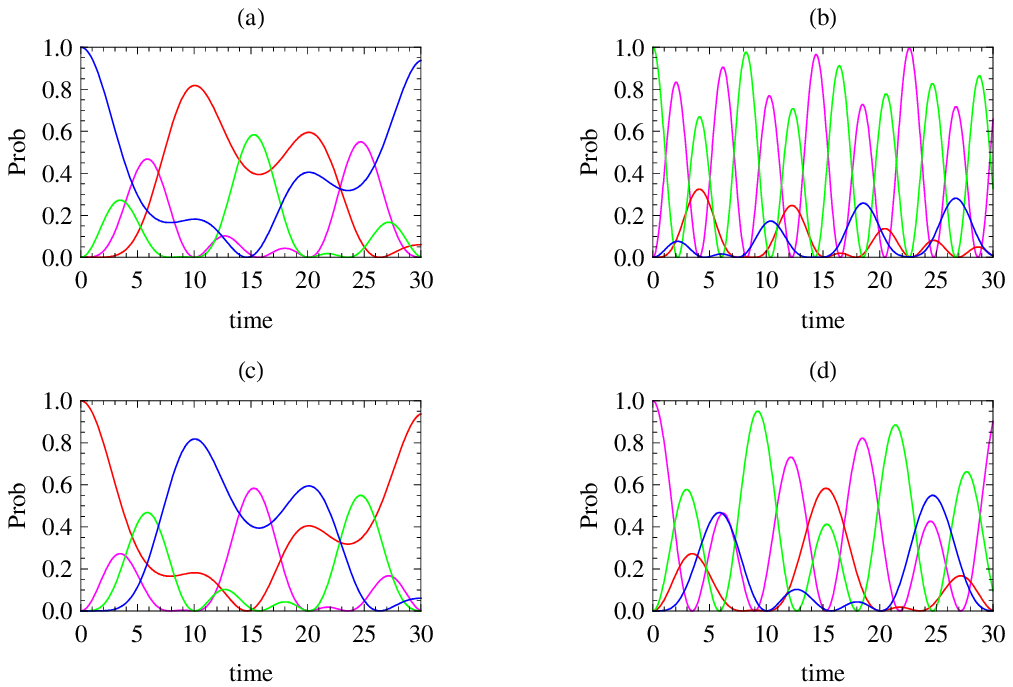}
\begin{flushleft}
Fig.11: Rabi oscillation of Model-V for Case-I (a), Case-II (b), Case-III (c) and Case-IV (d). Comparing with Fig.8 of Model-II, the inversion sysmmetry is clearly evident.
\end{flushleft}
\end{figure}
\begin{figure}[ht]
\includegraphics[width=8.25cm,height=8.25cm,keepaspectratio]{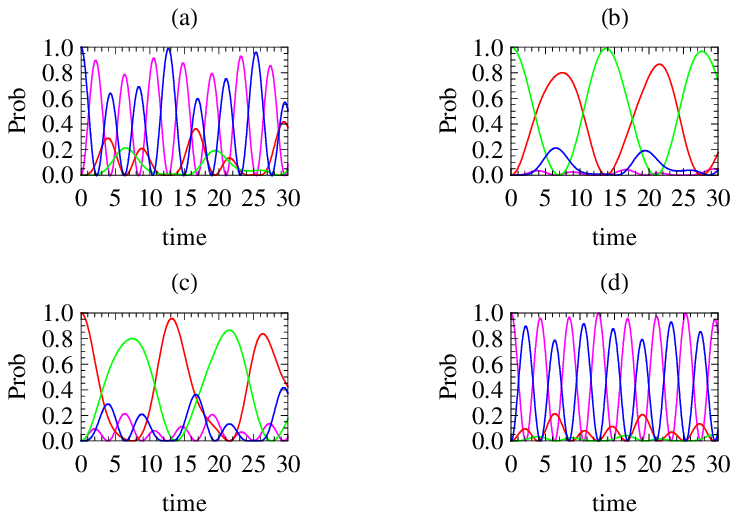}
\begin{flushleft}
Fig.12: Rabi oscillation of Model-VI for Case-I (a), Case-II (b), Case-III (c) and Case-IV (d). Comparing with Fig.7 of Model-I, the inversion sysmmetry is clearly evident.
\end{flushleft}
\end{figure}
and Case-IV: $C_1(0)=0, C_2(0)=0, C_3(0)=0, C_4(0)=1$, i.e., when the system is in the uppermost state labeled as level-4 (magenta), respectively.
To maximize the symmetry in the pattern of Rabi oscillation, we take the coupling parameters to be, $\kappa_{41}=.7$, $\kappa_{42}=\kappa_{31}=.4$ and $\kappa_{21}=\kappa_{32}=\kappa_{43}=.24$, respectively. Apart from that, at resonance, the detuning are taken to be $\Delta _{43}^I=0$, $\Delta _{32}^I=0$, $\Delta _{21}^I=0$, $\Delta _{31}^I=0$ and $\Delta _{42}^I=0$ for all models and the and the Rbi oscillation of various levels for all four cases are shown in Fig.7-12. If we compare the Rabi oscillation of Model-I (Model-II) shown in Fig.7 (Fig.8) with that of Model-VI (Model-II) shown in Fig.12 (Fig.10), an inversion symmetry is clearly evident. The existence of the inversion symmetry is a unique feature of the multilevel system which was also observed in the three-level cascade system~\cite{Nath2003,Nath2008b}. Furthermore, we note that for Model-III, the Rabi oscillation of Fig.9a is similar to Fig.9(d) and Fig.9(b) is similar to Fig.9(c), repectively. The existence of the inversion symmetry within same system has already been pointed out in the equidistant three-level cascade system~\cite{Nath2003,Nath2008b}. Finally in Model-III, taking the coupling parameters $\kappa_{43}\rightarrow \sqrt{3}\kappa_{43}$, $\kappa_{32}\rightarrow {2}\kappa_{32}$ and $\kappa_{21}\rightarrow \sqrt{3}\kappa_{21}$ to be in Eq.~(\ref{nine}) (or equivalently, in Eq.(A.6)), and the interacting field mode as monochromatic field, i.e., $\omega_{43}=\omega_{32}=\omega_{21}$, we recover the Hamiltonian as well as the Rabi oscillation of for the equidistant four-level system of spin-$\frac{3}{2}$ representation of $SU(2)$ symmetry~\cite{Nath2008a} indicating the consistency of our approach.
\section{\label{6x} Conclusion}
The primary objective of the paper is to present a detailed and systematic classification of the four-level system. To achieve this goal, we have discussed the tenets of $SU(4)$ group necessary to formulate the model Hamiltonians in terms of the shift operator of that group. We emphasize here that the selection rule allowed by the phenomenological tripod or inverted Y-type model studied as a representative model of four-level system studied in different context ~\cite{Qi2010, Li2008} are no longer an outcome of our classification. The exact solution of all six semi-classical models are obtained and the symmetry exists in the pattern of the Rabi oscillation of each model is illustrated. In a recent work, we have reported that the symmetric pattern of Rabi oscillation
is spontaneously broken on quantization of the cavity field not only for the cascade, lambda and vee systems~\cite{Nath2003,Nath2008b,Sen2012}, but also for the equidistant cascade four-level system~\cite{Nath2008a}. It is interesting to explore how the pattern of the symmetry is broken if we treat the tri-achromatic cavity field quantum mechanically. Finally we remark that it is reasonable to expect that above treatment can be generalized for $N$-level system which corresponds to $SU(N)$ group, however, we prefer to illustrate the exact solution of all possible configurations of the four-level system because it may form the basis of addressing a new class of coherent phenomena.
\begin{acknowledgements}
We thank Dr Tushar Kanti Dey, Dr Mihir Ranjan Nath and Dr Gautam Gangopadhyay for discussions. SS is thankful to Department of Science and Technology, New Delhi for partial support.
\end{acknowledgements}
\appendix*
\section{}
\par
In this Appendix we shall give the unitary operators which yields the time-dependent Hamiltonians and detuning frequency of the remaining models.
\par
The unitary operator which makes the Hamiltonian in Eq.(7) time-independent is given by,
\begin{eqnarray}
U_{II}(t) &=& exp[({\frac{i}{4}(-3{\omega _{21}} + 2{\omega _{31}} - 3{\omega _{41}})}{Z_3}\\ \nonumber
&+&{\frac{i}{2}(-2{\omega _{21}}-{\omega _{32}}+{\omega _{41}})}{T_3}\\ \nonumber
&+&\frac{i}{4}(-2{\omega _{21}}-3{\omega _{32}}+{\omega _{41}}){X_3})t],
\end{eqnarray}
and the time-independent Hamiltonian for Model-II takes the form,
\begin{equation}
{\tilde H_{II}(0)} =\left[ {\begin{array}{*{20}{c}}
{\Delta_{44}^{II}}&\kappa _{43}&0&0\\
\kappa _{43}&{\Delta_{33}^{II}}&0&\kappa _{31}\\
0&0&{\Delta_{22}^{II}}&{{\kappa _{21}}}\\
0&\kappa _{31}&{{\kappa _{21}}}&{\Delta_{11}^{II}}
\end{array}} \right],
\end{equation}
where the diagonal elements are defined as,
\begin{subequations}
\begin{eqnarray}
\Delta_{44}^{II}
&=& \frac{1}{4}( 3\Delta _{43}^{II} + 2 \Delta _{31}^{II} + \Delta _{21}^{II}),\\
\Delta_{33}^{II}
&=& \frac{1}{4}(2\Delta _{31}^{II} - \Delta _{21}^{II} - \Delta _{43}^{II}),\\
\Delta_{22}^{II}
&=& \frac{1}{4}(3\Delta _{21}^{II} - 2\Delta _{31}^{II} - \Delta _{43}^{II}),\\
\Delta_{11}^{II}
&=&  - \frac{1}{4}(2\Delta _{31}^{II} + \Delta _{43}^{II} + \Delta _{21}^{II}),
\end{eqnarray}
\end{subequations}
with the detuning frequencies given by,
\begin{subequations}
\begin{eqnarray}\label{eq:tenp1}
\Delta _{43}^{II} &=& ({\omega_2} - {\omega_3}) - {\omega _{43}}, \\
\Delta _{31}^{II} &=& (2{\omega _2} + {\omega_1} - \frac{1}{2}{\omega_3}) - {\omega_{31}},\\
\Delta _{21}^{II} &=& ({\omega_3} + 2{\omega_1}) - {\omega_{21}},
\end{eqnarray}
\end{subequations}
respectively.
\par
The unitary operator that makes the Hamiltonian in Eq.(8) time-independent is given by,
\begin{eqnarray}
U_{III}(t) &=& exp[({\frac{i}{4}(-3{\omega _{21}} - 2{\omega _{32}} - 3{\omega _{43}})}{Z_3}\\ \nonumber
&+&{\frac{i}{4}(-{\omega _{21}}-2 {\omega _{32}}+3 {\omega _{43}})}{T_3}\\ \nonumber
&+&\frac{i}{2}(-{\omega _{21}}-{\omega _{32}}-{\omega _{43}}){U_3})t].
\end{eqnarray}
Thus the time-independent Hamiltonian for Model-III is given by,
\begin{equation}
{\tilde H_{III}(0)} =\left[ {\begin{array}{*{20}{c}}
{\Delta_{44}^{III}}&\kappa _{43}&0&0\\
\kappa _{43}&{\Delta_{33}^{III}}&{{\kappa _{32}}}&0\\
0&{{\kappa _{32}}}&{\Delta_{22}^{III}}&{{\kappa _{21}}}\\
0&0&{{\kappa _{21}}}&{\Delta_{11}^{III}}
\end{array}} \right],
\end{equation}
where the diagonal elements are given by,
\begin{subequations}
\begin{eqnarray}
\Delta_{44}^{III}
&=& \frac{1}{4}( \Delta _{21}^{III} + 2 \Delta _{31}^{III} + 3\Delta _{43}^{III}),\\
\Delta_{33}^{III}
&=& \frac{1}{4}(\Delta _{21}^{III} - 2\Delta _{32}^{III} - \Delta _{43}^{III}),\\
\Delta_{22}^{III}
&=& \frac{1}{4}(\Delta _{21}^{III} - 2\Delta _{32}^{III} - \Delta _{43}^{III}),\\
\Delta_{11}^{III}
&=&  - \frac{1}{4}(3\Delta _{21}^{III} + 2\Delta _{32}^{III} + \Delta _{43}^{III}),
\end{eqnarray}
\end{subequations}
with the detuning three frequencies defined by,
\begin{subequations}
\begin{eqnarray}\label{eq:tenp1}
\Delta _{21}^{III} &=& (2{\omega _1} - {\omega_3}) - {\omega_{21}},\\
\Delta _{32}^{III} &=& (2{\omega_3} -{\omega_1} - {\omega_2}) - {\omega_{32}},\\
\Delta _{43}^{III} &=& (2{\omega_2} - {\omega_3}) - {\omega _{43}},
\end{eqnarray}
\end{subequations}
respectively.
\par
The unitary operator for the Hamiltonian in Eq.(9) is given by,
\begin{eqnarray}
U_{IV}(t) &=& exp[({\frac{i}{4}(-3{\omega _{21}} + 2{\omega _{41}} - {\omega_{43}})}{Z_3}\\ \nonumber
&+&{\frac{i}{2}(-{\omega _{21}} + {\omega _{41}} - 3 {\omega _{43}})}{T_3}\\ \nonumber
&+&\frac{i}{2}({\omega _{21}}-{\omega _{41}}-{\omega _{43}}){W_3})t],
\end{eqnarray}
and the time-independent Hamiltonian for Model-IV becomes,
\begin{equation}
{\tilde H_{IV}(0)} =\left[ {\begin{array}{*{20}{c}}
{\Delta_{44}^{IV}}&\kappa _{43}&0&\kappa _{41}\\
\kappa _{43}&{\Delta_{33}^{IV}}&0&0\\
0&0&{\Delta_{22}^{IV}}&\kappa _{21}\\
{{\kappa _{41}}}&0&\kappa _{21}&{\Delta_{11}^{IV}}
\end{array}} \right],
\end{equation}
where the diagonal elements are given by,
\begin{subequations}
\begin{eqnarray}
\Delta_{44}^{IV}
&=& \frac{1}{4}( -\Delta _{21}^{IV} + 2 \Delta _{41}^{IV} + \Delta _{43}^{IV}),\\
\Delta_{33}^{IV}
&=& \frac{1}{4}( -\Delta _{21}^{IV} + 2 \Delta _{41}^{IV} - 3 \Delta _{43}^{IV}),\\
\Delta_{22}^{IV}
&=& \frac{1}{4}(3\Delta _{21}^{IV} - 2 \Delta _{41}^{IV} + \Delta _{43}^{IV}),\\
\Delta_{11}^{IV}
&=&  - \frac{1}{4}(- \Delta_{21}^{IV} -2 \Delta_{41}^{IV} + \Delta _{43}^{IV}),
\end{eqnarray}
\end{subequations}
with the detuning three frequencies defined by,
\begin{subequations}
\begin{eqnarray}\label{eq:tenp1}
\Delta _{21}^{IV} &=& (2{\omega _1} + {\omega_3}) - {\omega_{21}},\\
\Delta _{41}^{IV} &=& ({\omega_1}+\frac{1}{2}{\omega_2}+2{\omega_3}) - {\omega_{41}},\\
\Delta _{43}^{IV} &=& ({\omega_2} + {\omega_3}) - {\omega _{43}},
\end{eqnarray}
\end{subequations}
respectively.
\par
The unitary operator for Hamiltonian in Eq.(10) is given by,
\begin{eqnarray}
U_{V}(t) &=& exp[({\frac{i}{4}(-3{\omega _{21}} - 2{\omega _{42}} - {\omega _{43}})}{Z_3}\\ \nonumber
&+&{\frac{i}{4}({\omega _{21}}+2 {\omega _{42}}-3 {\omega _{43}})}{T_3}\\ \nonumber
&+&\frac{i}{2}(-{\omega _{21}}-{\omega _{42}}+{\omega _{43}}){V_3})t].
\end{eqnarray}
and the time-independent Hamiltonian for Model-V is given by,
\begin{equation}
{\tilde H_V(0)} =\left[ {\begin{array}{*{20}{c}}
{\Delta_{44}^{V}}&\kappa _{43}&\kappa _{42}&0\\
\kappa _{43}&{\Delta_{33}^{V}}&0&0\\
\kappa _{42}&0&{\Delta_{22}^{V}}&\kappa _{21}\\
0&0&\kappa _{21}&{\Delta_{11}^{V}}
\end{array}} \right]
\end{equation}
where the diagonal elements are given by,
\begin{subequations}
\begin{eqnarray}
\Delta_{44}^{V}
&=& \frac{1}{4}( \Delta _{21}^{V} + 2 \Delta _{42}^{V} + \Delta _{43}^{V}),\\
\Delta_{33}^{V}
&=& \frac{1}{4}( \Delta _{21}^{V} + 2 \Delta _{42}^{V} - 3 \Delta _{43}^{V}),\\
\Delta_{22}^{V}
&=& \frac{1}{4}(\Delta _{21}^{V} - 2 \Delta _{42}^{V} + \Delta _{43}^{V}),\\
\Delta_{11}^{V}
&=&  - \frac{1}{4}(-3 \Delta_{21}^{V} -2 \Delta_{42}^{V} + \Delta _{43}^{V}),
\end{eqnarray}
\end{subequations}
with the detuning frequencies defined by,
\begin{subequations}
\begin{eqnarray}\label{eq:tenp1}
\Delta _{21}^{V} &=& (-{\omega_3} + {\omega_1}) - {\omega _{21}}, \\
\Delta _{42}^{V} &=& ({\omega_2}-{\omega_1}+2{\omega_3}) - {\omega_{42}},\\
\Delta _{43}^{V} &=& ({\omega_3} + 2{\omega_2}) - {\omega_{43}},
\end{eqnarray}
\end{subequations}
respectively.
\par
The unitary operator that makes the Hamiltonian in Eq.(11) time-independent is given by,
\begin{eqnarray}
U_{VI}(t) &=& exp[({\frac{i}{4}({\omega _{32}} - 3{\omega _{41}} - 2{\omega _{43}})}{W_3} \\ \nonumber
&+&{(-{\omega _{32}}+2 {\omega _{41}}-2 {\omega _{43}})}{T_3}\\ \nonumber
&+&\frac{i}{4}(-3{\omega _{32}}+{\omega _{41}}-2{\omega _{43}}){U_3})t],
\end{eqnarray}
and the time-independent Hamiltonian for Model-VI is given by,\\
\begin{equation}
{\tilde H_{VI}(0)} =\left[ {\begin{array}{*{20}{c}}
{\Delta_{44}^{VI}}&\kappa _{43}&0&\kappa _{41}\\
\kappa _{43}&{\Delta_{33}^{VI}}&\kappa _{32}&0\\
0&{{\kappa _{32}}}&{\Delta_{22}^{VI}}&0\\
\kappa _{41}&0&0&{\Delta_{11}^{VI}}
\end{array}} \right]
\end{equation}
where the diagonal elements are given by,
\begin{subequations}
\begin{eqnarray}
\Delta_{44}^{VI}
&=& \frac{1}{4}( \Delta _{32}^{VI} + \Delta _{41}^{VI} + 2\Delta _{43}^{VI}),\\
\Delta_{33}^{VI}
&=& \frac{1}{4}( \Delta _{32}^{VI} + \Delta _{41}^{VI} - 2 \Delta _{43}^{VI}),\\
\Delta_{22}^{VI}
&=& \frac{1}{4}(-3\Delta _{32}^{VI} + \Delta _{41}^{VI} -2 \Delta _{43}^{VI}),\\
\Delta_{11}^{VI}
&=&  - \frac{1}{4}(-\Delta_{32}^{VI} -3 \Delta_{41}^{VI} + 2\Delta _{43}^{VI}),
\end{eqnarray}
\end{subequations}
with the detuning frequencies defined by,
\begin{subequations}
\begin{align}
\Delta _{43}^{VI} &= ({\omega_2}-{\omega_3} + {\omega_1}) - {\omega _{43}}, \\
\Delta _{32}^{VI} &= (2{\omega_3} - \frac{1}{2} {\omega_2}) - {\omega_{32}},\\
\Delta _{41}^{VI} &= (\frac{1}{2}{\omega_2}-2{\omega_1}) - {\omega_{41}},
\end{align}
\end{subequations}
respectively.
\bibliography{fourbib}
\vfill
\end{document}